\documentclass[preprint]{JASA}

\begin{document}
%% the square bracket argument will send term to running head in
%% preprint, or running foot in reprint style.

\title[]{Supersonic and Superluminal Energy and Speed of Information via Temporal Interference in a Dispersionless Environment}

% ie
%\title[JASA/Sample JASA Article]{Article title should be less than 17 words, no acronyms}

%% repeat as needed

\author{John L. Spiesberger}
\email{john.spiesberger@gmail.com}
\affiliation{Dept. of Earth and Environmental Science, U. of Pennsylvania, Philadelphia, PA 19104, USA}
\author{Eugene Terray}
\email{eterray@whoi.edu}
\affiliation{Woods Hole Oceanographic Institution}

% ie
%\author{Author One}
%\author{Author Two}
%\author{Author Three}

%\affiliation{}

% ie
%\affiliation{Department1,  University1, City, State ZipCode, Country}

%\altaffiliation{}

% may be added after \author{}, ie
% \altaffiliation{Also at: Department1,  University1, City, State ZipCode, Country.}

%% for corresponding author
\email{john.spiesberger@gmail.com}

%% For preprint only,
%  optional, if you want want this message to appear in upper right corner of title page
% \preprint{}

%ie
%\preprint{Author, JASA}		

% optional, if desired:
%\date{\today} 

\begin{abstract}

Numerical implementation of a theory yields acoustic wave packets whose peak-to-peak speeds, $c_{3d}$, are supersonic in a
dispersionless medium due to temporal  interference between direct and boundary-reflected paths.
The effect occurs when the source and receiver are near each other and at least one is within $c\tilde{\delta t}/2$
of the boundary, where $c$ is the phase speed of propagation in the medium, and $\tilde{\delta t}$ 
is the smallest temporal separation between the paths at which interference first occurs.
This direct+reflected path effect is distinct from previously-observed superluminal phenomena and theories including
quantum tunneling, cavity vacuum fluctuations,  and  group speeds due to anomalous dispersion.
For temporally interfering direct+reflected paths, simulations yield a speed of information  less than $c$.
The speed of information from the interfering paths can exceed the speed  derived from
propagation  only along the direct path.
We conjecture these results will also hold for electromagnetic (EM) wave propagation. If so, we prove the speed of information
is less than or equal to the  speed of light in a vacuum, so the effect
does not violate special relativity.
These theoretical and simulation results, as well as their conjectured EM extension, should be readily accessible to experimental verification.
\end{abstract}

%% pacs numbers not used

\maketitle

%  End of title page for Preprint option --------------------------------- %

%% See preprint.tex/.pdf or reprint.tex/.pdf for many examples

\section{\label{sec:1} Introduction}

\begin{figure}[ht]
\centerline{\includegraphics[width=6in]{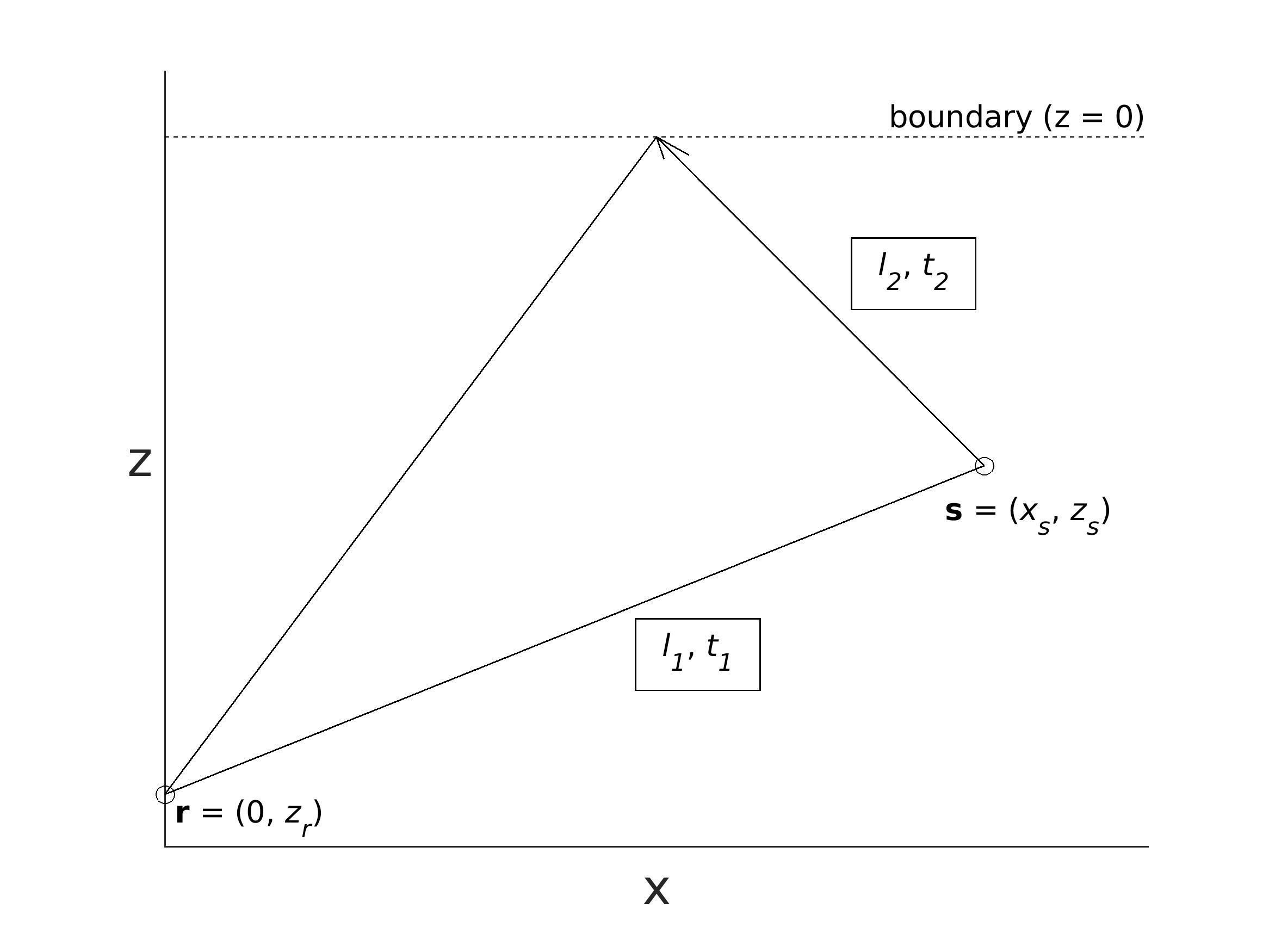}}
\caption{\label{fig:definition_fig} Signal propagates from source ${\bf s}$ to receiver ${\bf r}$ along direct and boundary reflected paths
  with lengths,  $l_1$ and $l_2$,  and propagation times, $t_1$  and $t_2$, respectively. The $y$ axis is not shown but the $x-y$ plane is perpendicular to the
$z$ axis and the boundary is in a $x-y$ plane.}
\end{figure}

Recent theoretical results for dispersionless media predict acoustic wave packets can occur at locations with speeds significantly slower or faster than the
phase speed when the source and receiver are near each other, are near a reflecting boundary, and 
the waveforms from the direct and reflected paths temporally interfere \cite{c3d_slow,c3d_fast} (Fig. \ref{fig:definition_fig}).
In other words, a wave packets' speed is modified even when the phase and group speeds are identical. 
For the direct+reflected path effect,  the so-called $c_{3d}$ speed of a  wave packet is defined as,
\begin{equation}
c_{3d} \equiv l_1/t_m \ , \label{eq:c3d}
\end{equation}
where $l_1$ is the
distance of the straight  path from source to receiver, and $t_m$ is the time between the wave packets. The $c_{3d}$
can be unequal to the group speed.  Unless noted otherwise, $t_m$ is the interval between the first peak of the
absolute value of  the Hilbert transform of the transmitted and received time
series.
This slowing and speeding of wave packets may need to be accounted for
when deriving correct confidence intervals of location based on measurements of propagation time or measurements of the time-differences of arrival
(TDOA) between receivers \cite{c3d_slow,c3d_fast}.  The physics of the direct+reflected path effect
only requires a reflecting boundary with a nearby source or receiver, suggesting the same  phenomenon
might occur for electromagnetic (EM) waves. Because the effect yields supersonic values of the $c_{3d}$ in dispersionless media,
we  conjecture it could yield superluminal values of the $c_{3d}$ for EM  waves propagating in a vacuum near a reflecting boundary.

% Figure fromm figures/definition_figure/rays_str_reflected_receiver_and_source2.jpg

It has long been known superluminal group speeds occur due to  anomalous dispersion \cite{leroux,diener,Wang2000}, where the frequencies of an EM  wave  packet
interact with ions  near  their resonant frequencies.  In order to  preserve the idea of causality in
special relativity \cite{einstein}, the second postulate was modified to mean  the speed of information  cannot exceed the
speed of light in a vacuum
\cite{sommerfeld,brillouin,brillouin_translated_papers,diener}. Scientists also observed superluminal phenomena due to 
microwave tunneling \cite{enders_nimitz_1992,PhysRevE.62.5758},  quantum tunneling \cite{chiao_and_steinberg_1997,PhysRevLett.71.708}, and
discussed the possibility
in the context of
quantum electrodynamics \cite{SCHARNHORST1990354}. None of these investigations yielded 
a speed of information exceeding the speed of light in a  vacuum.

Consequently, the speed of information due to the direct+reflected path effect is simulated in this paper for acoustic waves
following methods from information theory employed by \citet{stenner_2003}. Whether \citet{stenner_2003} measured superluminal energy
was discussed  shortly thereafter \cite{nimtz2004superluminal}.

\citet{Robertson2007Soundbeyond} indirectly inferred the existence of superluminal propagation of acoustic pulses in a short loop tube
attached to a long tube.  The loop tube introduced other acoustic paths, destructively and constructively interfering with the otherwise
undisturbed signal at certain
resonant frequencies.  This anomalous dispersion, and caused the group speed to increase.
An acoustic speaker broadcasted narrowband acoustic waves, near a resonant frequency of the loop,
at one end of a straight  tube of length eight meters.
The propagation time was derived from data collected at a  microphone at the other end of the eight meter tube.
Times of the arriving acoustic energy were made with and without the loop.
The arriving energy packet was  advanced by 0.0024 s when the loop was  present. Assuming a sound speed in air of 330 m/s, the speed of the
energy packet in the absence of resonances is about $ 8 \ \mbox{m} / 330 \ \mbox{(m/s)} \sim 0.024$ s, 
Thus, the propagation time  of the wave packet with the loop is
0.024 - 0.0024=0.0218 s, yielding a measured group speed of $8 \ \mbox{m} / 0.0218 \ \mbox{s} \sim  367$ m/s, much less than the speed of light in a vacuum.
However, they explain there must be superluminal speeds in the proximity of the  loop to cause this advance.
They did not attempt to directly measure superluminal  speeds because they believed placing the speaker and  microphone directly
at the start and end of the loop would destroy the natural resonances of acoustic waves in the loop and suppress the superluminal effect.
Interestingly, \citet{Robertson2007Soundbeyond}  state their phenomenon is like the so-called ``Comb'' effect from the field
of architectural acoustics, wherein sound arriving at a listener is a combination of the direct path and one reflecting from a hard boundary, causing
destructive and constructive interference at the listener's head at certain frequencies, and degrading the music's fidelity \cite{comb_filt}.  
The direct+reflected path effect seems to be  the next step, recognizing interference of direct and reflected paths
modifies the  $c_{3d}$, i.e.   the speed of appearance of acoustic wave packets.

\section{\label{sec:wave_eqn_soln} Exact solutions for acoustic  waves}

\citet{c3d_slow} recalled the exact three-dimensional solutions to the linear acoustic wave equation in the presence of an ideal flat boundary,
\begin{eqnarray}
\rho_a(x,y,z,t) &=& a_1 s(t-l_1) - a_2 s(t-l_2) \ , \label{eq:ac_soln_0_press} \\
\rho_b(x,y,z,t) &=& a_1 s(t-l_1) + a_2 s(t-l_2)) \ , \label{eq:ac_soln_0_vel} \ 
\end{eqnarray}
where $\rho_a(x,y,z,t)$ is for a boundary with zero fluctuations of pressure and $\rho_b(x,y,z,t)$ is for zero normal velocity.
When energy spreads spherically  from the sources, $a_i = 1/l_i$, where the  distances,
$l_1$ and $l_2$ are $l_1=\sqrt{(x_s-x_r)^2 + (y_s-y_r)^2 + (z_s-z_r)^2}$ and $l_2=\sqrt{(x_s-x_r)^2 + (y_s-y_r)^2 + (z_s+z_r)^2}$ with
source and receiver at $(x_s,y_s,z_s)$ and $(x_r,y_r,z_r)$ respectively in Cartesian coordinates (Fig. \ref{fig:definition_fig}). These solutions
are most easily understood by introducing an image source on the opposite side of the boundary. It is $180^{\circ}$ out of phase for the boundary with zero pressure
fluctuation and in phase for the boundary with zero normal velocity.  $s(t)$ is the emitted waveform.

\section{\label{speed_information} Speed of Information}

\citet{stenner_2003} state they measured the speed of information for EM waves where the group speed exceeded the speed of light in a vacuum.  They
transmitted two symbols and found through experiment the speed of information was less than the speed of light in a vacuum. Their procedures
are duplicated  here to theoretically derive the speed of information  when the $c_{3d}$ exceeds both the phase and group speeds, $c$,
due to temporal  interference between
the direct and reflected paths.

\begin{figure}[ht]
\centerline{\includegraphics[width=6in]{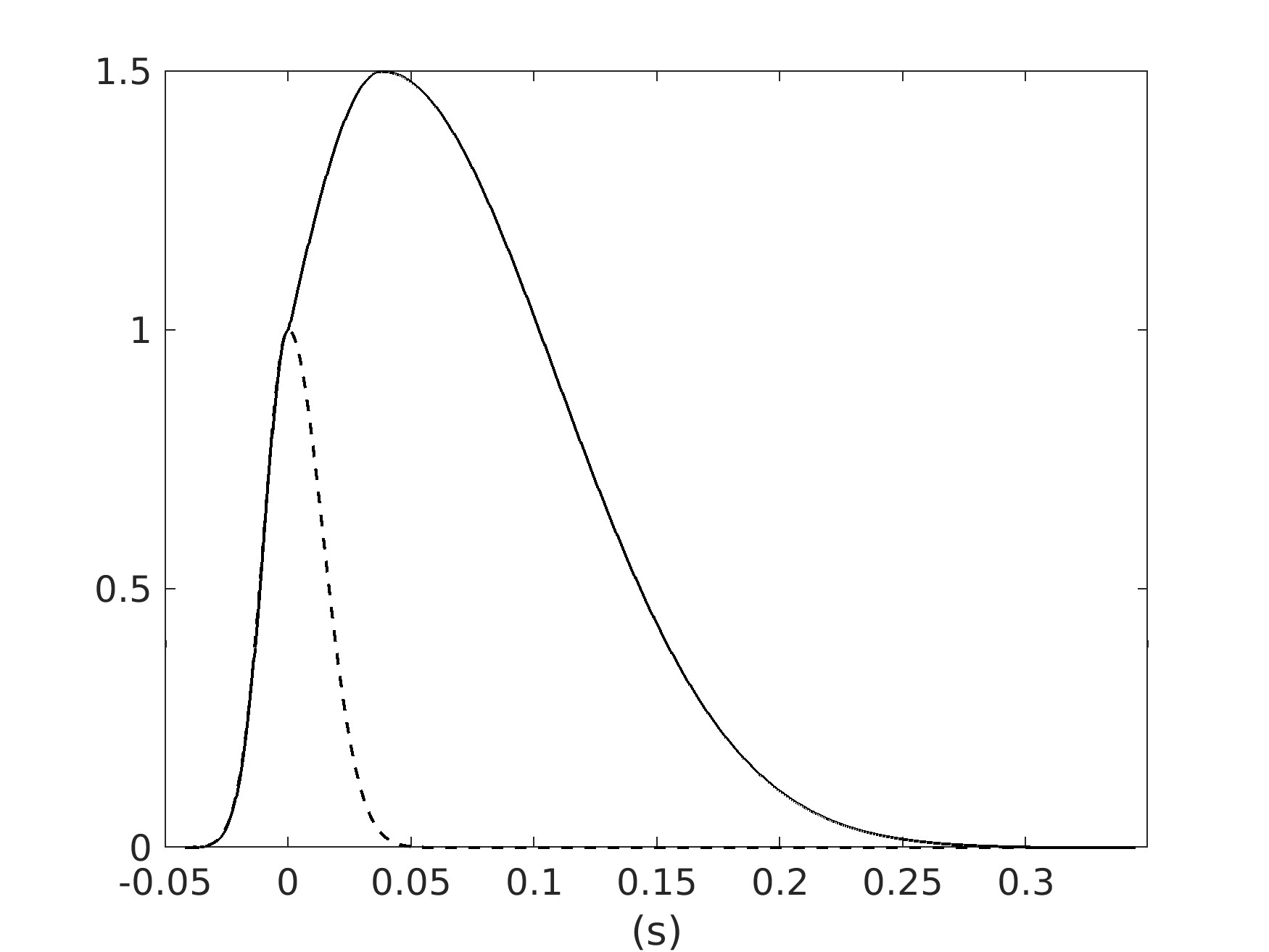}}
\caption{\label{fig:ideal_symbols_2} Envelopes of ideal symbols following Eqs. \ref{eq:ideal_before_0} through \ref{eq:t_max}.
  Point of non-analyticity is time zero.
Symbols 0 and 1 differ only at times exceeding zero with symbol 0 being dashed.}.
\end{figure}

Symbols 0 and 1 are detected at the receiver, where they are identical up to time $t=0$, defined to be the time when a switch is
thrown to choose symbol 0 or 1 thereafter (Fig. \ref{fig:ideal_symbols_2}).  The point of non-analyticity occurs at $t=0$, and, in our formulation,
there is a discontinuity in the first derivative of the ideal transmitted waveform.
Both symbols have  carrier frequency, $f_s$, but are distinguished
by different envelopes.  Before the switch is thrown, the waveform is,
\begin{equation}
w_a(t) = \cos(2 \pi f_s t) \exp[-(t/\tau_a)^2 ] \ ; \ t \leq 0 \ . \label{eq:ideal_before_0}
\end{equation}
The envelope for symbol 0 goes down for $t > 0$ as,
\begin{equation}
w_0(t) = \cos(2 \pi f_s t) \exp[-(t/\tau_b)^2 ] \ ; \ t > 0 \ . \label{eq:ideal_symb_0}
\end{equation}
The envelope for symbol 1 goes up at $t=0$  to maximum amplitude, $a_{max}$, at $t=t_{max}$ and then down afterwards as,
\begin{eqnarray}
  w_1(t) &=& a_{max} \cos(2 \pi f_s t) \exp[ -((t-t_{max})/\tau_c)^2 ] \ ; \ 0 < t \leq t_{max} \label{eq:ideal_symb_1_up} \\
  w_1(t) &=& a_{max} \cos(2 \pi f_s t) \exp[ -((t-t_{max})/\tau_d)^2 ]\ ; \ t_{max} < t \ . \label{eq:ideal_symb_1_down} 
\end{eqnarray}
where
\begin{equation}
t_{\max} = \tau_b \sqrt{\ln(a_{max})} \ . \label{eq:t_max} 
\end{equation}
See Fig. \ref{fig:ideal_symbols_2} for an  example.

% Is ideal_symbols_2.jpg

\subsection{\label{sect:symbol_detection} Symbol classification}

The speed of information is measured from the time the switch is thrown to the time the symbol is reliably detected.  Let $R_j(t)$ be the time series
at the receiver
for symbols $j=0 \ \mbox{and} \ 1$ in the absence of noise.  The symbols can be classified without error at the least time they differ, i.e. $T_{differ}$.
This time decreases with bandwidth, and vanishes for infinite bandwidth filters.

To mimic reality, the switch and the receiver's detector have finite bandwidth, each with their own time delay.
The bandwidth's are made large to minimize their delays.  Following \citet{stenner_2003} and principles of signal detection (pp. 88-94 of  \citet{Madhow_2008}),
symbol classification at finite SNR becomes a
statistical problem employing the decision statistic,
\begin{equation}
D(\tau, \chi) \equiv {L}_0(\tau,\chi) - {L}_1(\tau,\chi) \ , \label{eq:D}
\end{equation}
where $L_j(\tau,\chi)$ are outputs of a matched filter operating on the received time series,
\begin{equation}
  \chi_j(t) = R_j(t) + n(t) \ , \label{eq:chi0}
\end{equation}
where $n(t)$ is noise, and the matched filters for symbols 0 and 1 are,
\begin{eqnarray}
  L_0(\tau,\chi) &=& \int_{t_s}^{t_s+\tau} \chi_j(t) R_0(t) dt / [\alpha_0(\tau_{\alpha})  N_0(\tau)] \ , \label{eq:L0_0} \\
  L_1(\tau,\chi) &=& \int_{t_s}^{t_s+\tau} \chi_j(t) R_1(t) dt / [\alpha_1(\tau_{\alpha})  N_1(\tau)] \ . \label{eq:L0_1}  
\end{eqnarray}
The denominators are normalizing terms defined as,
\begin{eqnarray}
  \alpha_0(\tau_{\alpha}) &\equiv& \int_{t_s}^{t_s+\tau_{\alpha}} \chi^2_0(t) / N_0(\tau_{\alpha}) \ , \label{eq:alpha0} \\
  N_0(\tau) &\equiv& \int_{t_s}^{t_s + \tau} R_0^2 dt \ ,  \label{eq:N0} \\
  \alpha_1(\tau_{\alpha}) &\equiv& \int_{t_s}^{t_s+\tau_{\alpha}} \chi^2_1(t) / N_1(\tau_{\alpha}) \ , \label{eq:alpha1} \\
  N_1(\tau) &\equiv& \int_{t_s}^{t_s + \tau} R_1^2 dt \ .  \label{eq:N1} 
\end{eqnarray}
The matched-filter is integrated starting at time $t_s$,  chosen to occur before the point of non-analyticity occurs at the receiver, so as to  accumulate
its energy between it and the ending evaluation time at $t_s+\tau$, where $\tau \geq t_s$ and continues on until the energy of the
time series has passed.  $\tau_{\alpha}$ is chosen to be just before the  time when the symbols separate.
For experiments, the replicas, $R_j(t)$, are estimated by averaging over many realizations to suppress noise. In this paper, they are
equal to the received waveforms without noise.

To estimate the bit error rate (BER), the probability density function of $D(\tau,\chi)$ is empirically derived for two scenarios.
In the  first, it is empirically estimated
from many noisy realizations of only symbol 0; i.e. many realizations of $\chi_0(t)$.  The second probability distribution of $D(\tau,\chi)$ is
derived from many noisy realizations of only symbol 1.
Each probability density function is fitted to a Gaussian distribution, and normalized to area of one-half.  Their area of
overlap is the BER \cite{stenner_2003}.

We verified our implementation of the BER empirically as follows.  First, the noiseless values for $D(\tau,\chi)$, denoted $\hat{D}_0(\tau,\chi_0)$ and $\hat{D}_1(\tau,\chi_1)$ for symbols 0 and 1,
are constructed separately for each symbol using the noiseless timeseries at the receiver. The hat indicates evaluation with no noise, i.e. $n(t)=0$.
Then, for any incoming noisy symbol at time $\tau$, $D(\tau,\chi)$ is computed.
Symbol 0 is declared to be received when $D(\tau)$ is closer to $\hat{D}_0(\tau,\chi_0)$.  Otherwise symbol 1 is declared received.
The BER rate from this procedure agrees with the procedure outlined by \citet{stenner_2003}.

\subsection{\label{simulation_1} BER example}
 
Consider an ideal scenario in water where the phase speed, $c$, is 1500 m/s and a compact source and receiver are located at Cartesian $(x,y,z)$ coordinates
(0,  0,  -11.746) m and (12.698, 0,  -41.995) m, respectively. 
The carrier and sample frequencies are 3948.34 Hz  and 100 kHz respectively.
For the ideal signal, $\tau_a$, $\tau_b$, $\tau_c$ and $\tau_d$ are 0.014, 0.02, and 0.06, and 0.1 s respectively (Fig. \ref{fig:ideal_symbols_2}).
Suppose the ideal emitted waveform is subject to an
elliptic filter with stop and passbands of [3800, 4200] and [3900 , 4100] Hz respectively, a stopband
attenuation of 30 dB, and a passband ripple of 1 dB. Also suppose the filter of the receiver's detector is the same.
The peak SNR at  the receivers is 50 dB.

With $t=0$, corresponding to the time the switch is thrown, the ``ideal'' speed of information is the phase speed, $c$ (Fig. \ref{fig:definition_fig}).
For acoustic waves in our scenarios, it equals the phase and group speeds, $c$. The ``ideal'' time of information is $t_1$,
and is the time the point of non-analyticity is received if the speed of information is $c$.
For application to EM waves (Sec. \ref{sec:sound_speed_information}),
this is the time to beat if the assumption of causality is overturned through observation.

\begin{figure}[ht]
\centerline{\includegraphics[width=6in]{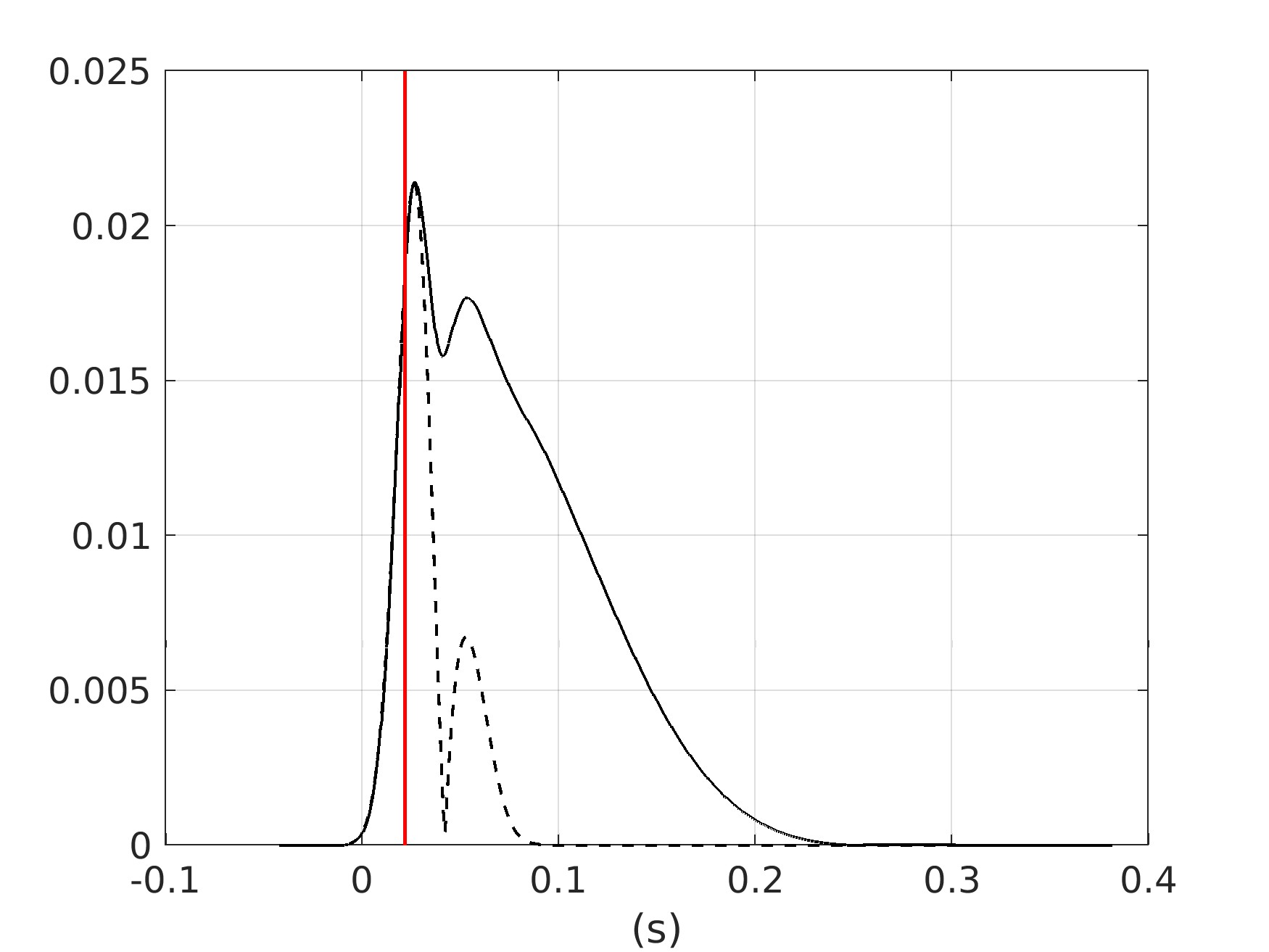}}
\caption{\label{fig:received_symbols_2} Envelopes of received symbol 0 (dashed) and 1 (solid). Red line is time the point of non-analyticity
  arrives at receiver in absence of any finite bandwidth effects along the direct path.}.
\end{figure}

\begin{figure}[ht]
\centerline{\includegraphics[width=6in]{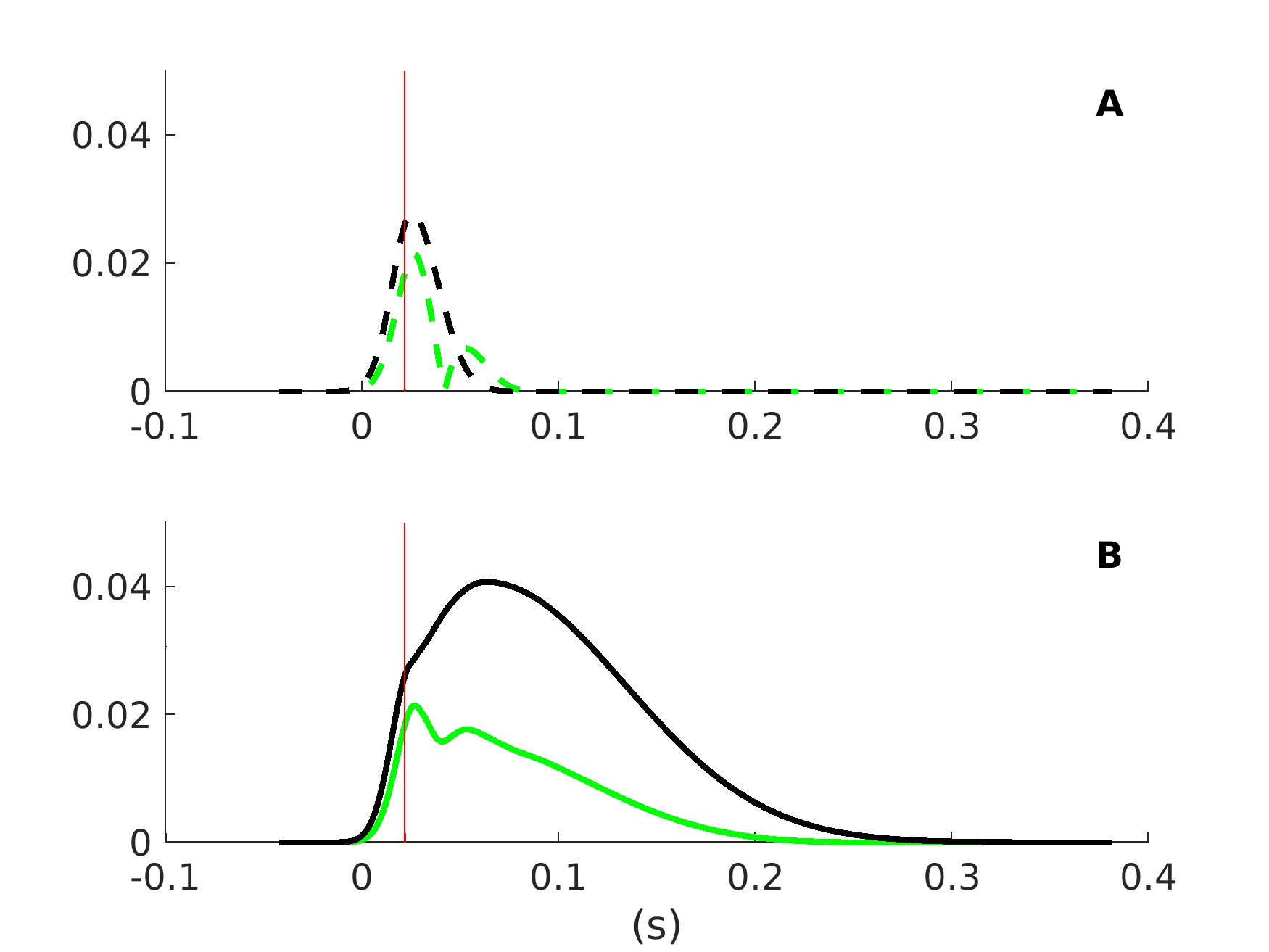}}
\caption{\label{fig:direct_vrs_interferring_paths_2} A)  Envelope of wave packet for symbol 0 at receiver for direct (black) and temporally interfering paths (green).
  Red line indicates time when point of non-analyticity arrives if energy propagates at phase and group speed, $c$, in absence of delays from
  finite-bandwidth effects at source and receiver. B) Same except for symbol 1.}.
\end{figure}

The flat boundary is assumed to have zero pressure fluctuations, and exact solutions are obtained from Eq. \ref{eq:ac_soln_0_press}.
The envelopes of the received symbols are derived by taking the absolute value of their Hilbert transforms, both
exhibiting their largest value at a time exceeding $t_1$ (Fig. \ref{fig:received_symbols_2}).  As expected, the envelopes
of the signals from  only the direct paths  differ from their temporally-interfering paths and the envelopes all differ 
after $t_1$ (Fig. \ref{fig:direct_vrs_interferring_paths_2}).

% Is received_symbols_2.jpg

% Is direct_vrs_interferring_paths_2.jpg

\begin{figure}[ht]
\centerline{\includegraphics[width=6in]{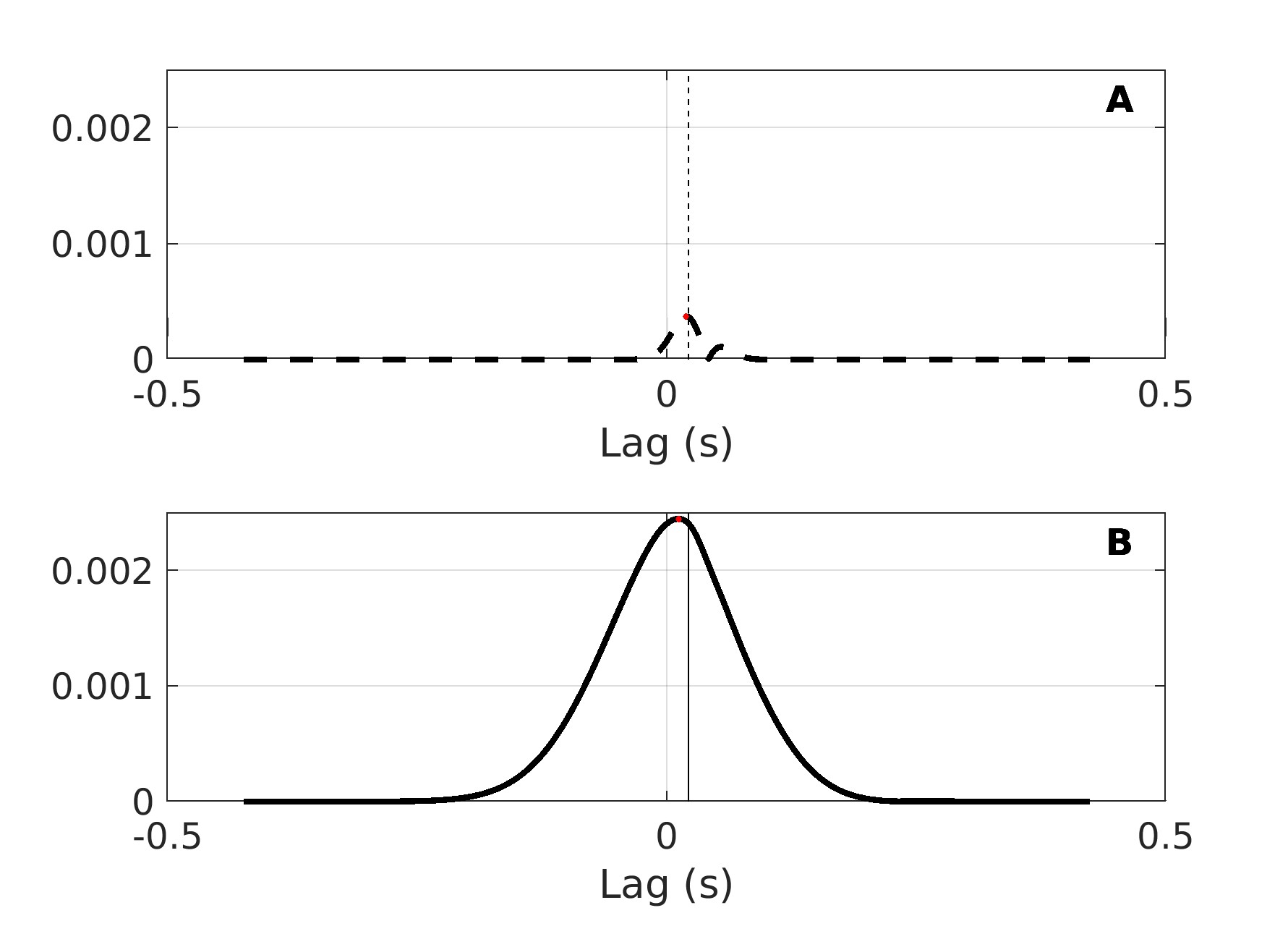}}
\caption{\label{fig:ccf_symbols_2} A)  Absolute value of Hilbert  transform of cross-correlation between emitted and received  symbol 0.
  Ideal time of arrival,
  $t_1$ is vertical dashed line. Red dot at peak.  B) Same except for symbol 1 and $t_1$ at vertical solid line.}.
\end{figure}

The $c_{3d}$ are derived by cross-correlating their emitted and received timeseries, taking the absolute value
of the Hilbert transform, and identifying the lag at the maximum peak, yielding 1694.5 and 2782.5 m/s for symbols 0 and 1 respectively (Fig. \ref{fig:ccf_symbols_2}).
Both exceed $c$, thus exhibiting modification of the occurrence of an energy envelope appearing at supersonic speeds in a dispersionless media.

% Is ccf_symbols_2.jpg

\begin{figure}[ht]
\centerline{\includegraphics[width=6in]{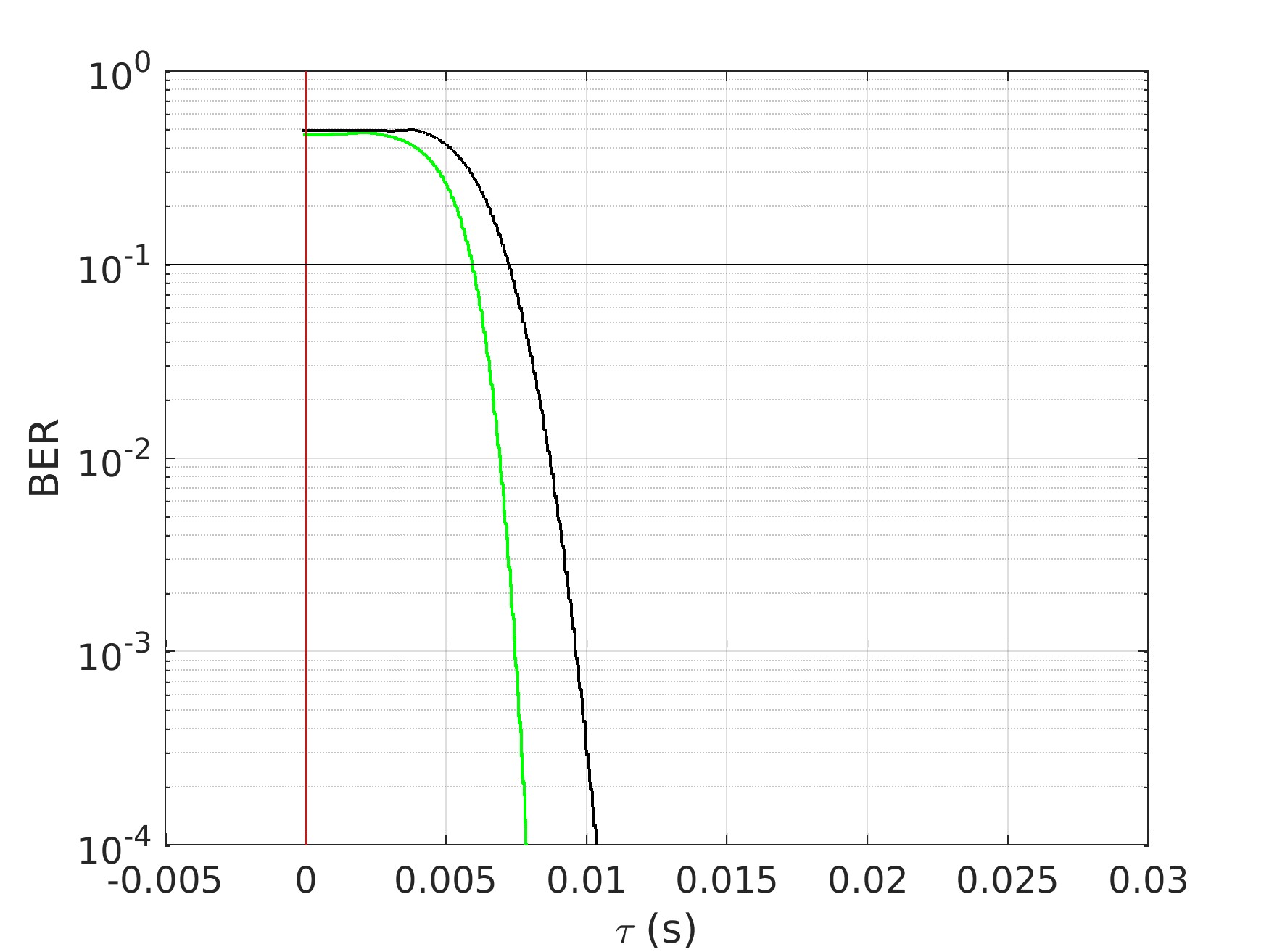}}
\caption{\label{fig:ber_2} BER for symbols 0 \& 1 assuming propagation occurs only along direct path (black) or temporally interfering paths (green).
  $tau$ axis equals zero at time series sample just before the point of non-analyticity arrives.
  Red line indicates value of $\tau$ if propagation occurs at phase and group speeds, $c$ and for infinite bandwidths of
the switch at the source and the receiver apparatus. BER of 0.1 indicated.}.

\end{figure}

For this example, the BER drops to 0.1 at about 0.005925 s for the temporally-interfering paths (Fig. \ref{fig:ber_2}). If the
simulation is conducted without a reflecting boundary, only the direct path arrives and the BER drops to 0.1 at 0.0071966 s (Fig. \ref{fig:ber_2}).
These both occur after $t_1$, so the speed of information is less than the phase and group speeds, $c$. Although the peak of the wave packet occurs
at supersonic speed, information is transmitted at subsonic speed.  Perhaps surprisingly, the speed of information for temporally interfering
paths exceeds the speed of information for the direct path only, andd both are less than $c$.

% Is ber_2.jpg

\section{\label{discussion} Discussion}

\subsection{\label{sec:sound_speed_information} Sound waves and speed of energy and information}

Acoustic energy and information can be conveyed at supersonic speeds via shock waves
\cite{lighthill1978waves}.  The absolute value of the Hilbert transform of an acoustic wave packet, derived from  the
direct+reflected path theory, is a measure of the square root of the energy of sound as a function of time, and thus a  measure
of energy as a function of time. Therefore, the simulations of the direct+reflected path  theory
do predict the appearance of supersonic speeds of energy packets but subsonic speed of information in the few simulations
conducted so far (Fig. \ref{fig:ber_2}).  This occurs in a dispersionless media.

Surprisingly, the  simulations predict the speed of information for the direct+reflected path effect can exceed the speed for the case with no reflected path at all
(Fig. \ref{fig:ber_2}). This phenomenon was always observed for the
few cases we simulated (not shown).  This bears further investigation.

\subsection{\label{sec:relativity} Special relativity and the speed of information}

The direct+reflected  path effect seems to be applicable to waves in general, so we conjecture the appearance of superluminal wave packets may exist
for EM waves. An experiment is needed to test
this conjecture.

\citet{sommerfeld} and \citet{brillouin,brillouin_translated_papers} explain the speed of a sharp discontinuous wavefront of an EM wave
moves through media at the speed of light
in a vacuum. This sharp wavefront is the so-called point of non-analyticity in more contemporary thinking, and can be replaced by any discontinuity
in the emitted waveform.
At the time of wavefront passage, they state ions interact with the EM wave, and their re-radiation leads to energy packets with superluminal speeds, but the
information flows with the discontinuous wavefront and re-radiation occurs later and therefore cannot increase the speed of information.  The experiment
conducted by \citet{stenner_2003} did not find any violation of this interpretation. 
Next, we prove the speed of information is less than or equal to the speed of light in a vacuum if the direct+reflected path effect is applicable to
EM radiation.

The reflected path's length exceeds the direct path's length. Consequently, with infinite
bandwidth filters,  the moment of non-analyticity in the time series has
infinite bandwidth
and zero temporal resolution, making it impossible for the direct and reflected paths to temporally
interfere.   The speed of the first arriving point of non-analyticity from the direct path travels at speed $c$, and no faster.

Consider a case with infinite SNR and infinite bandwidth filters. Further assume the sample interval is less than the difference in propagation time along
the reflected  and direct paths, i.e.   $t_2-t_1$, in Fig. \ref{fig:definition_fig}. Suppose the first a/d sample at  the receiver  is taken at the
same  time  the switch is thrown at the source selecting symbol 0 or 1.  Let $q$ denote the first a/d sample number at the receiver where the time series
from the interferring paths changes depending on whether symbol 0 or 1 is transmitted.  This distinguishing sample must be due only to transmission
along the direct path because the direct path arrives first and contributions of the reflected path for samples 1 through $q$ are the same regardless of the
transmitted symbol.

Now replace infinite with finite bandwidth filters, still
with infinite SNR. 
For real-time detection of information, causal filters must be used and the only samples available to distinguish the symbols up to sample number $q$ come
from samples 1 through $q$. Thus, the first sample number, $p$,  distinguishing
symbol 0 from 1 from the  direct+reflecting paths must obey $p\geq  q$, regardless of the details of their interference. The speed of information
cannot increase when going from  infinite to finite SNR.
Since the time corresponding to $q$ occurs at time $t_1 = c  l_1$  (Fig. \ref{fig:definition_fig}), the speed of information must be $c$ or less.
If the direct+reflecting path effect can be made to work with EM waves, the special theory of relativity is still valid.

\section{\label{conclusion} Conclusion}

It is not known if experimental verification of the theory of the direct+reflected path effect would be easier to conduct with acoustic or EM  waves,
but experimental verification is needed.
A beam splitter might send one beam toward the reflector
and the other toward the receiver, where they interfere. 
If lasers are used to explore the
direct+reflected  path effect,
the amplitudes of the direct and  reflected paths would not decay following the spherical spreading of energy, so $a_i=1$ in Eq. \ref{eq:ac_soln_0_press}, giving
more weight to interference from the reflected path.

We do not understand why the direct+reflected path effect yields a faster speed of information for temporally interfering paths than the speed derived in the  absence
of the reflected path, where the wave only propagates directly from source to receiver.  This may not be true for all types of symbols, and has only been
simulated with a few cases.  One possibility is interference emphasizes the differences between
the symbols just after the point of non-analyticity compared with propagation only along the direct path.

The direct+reflected path effect should yield wavepackets with supersonic speeds without a sonic boom. The former effect must transmit information
slower than the phase speed, $c$, in  a dispersionless media (Sec. \ref{sec:relativity}), but the sonic boom can transmit information at a  speed exceeding $c$.

%% before appendix (optional) and bibliography:
\begin{acknowledgments}

Research  was supported by Office of Naval Research grant N00014-23-1-2336.
\end{acknowledgments}

\section{\label{declarations} Author Declarations}
No conflicts of interest.

The data that support the findings of this study are available within the article.

% ---------------------------------------------------------------------
% Appendix  (optional)

%\appendix
%\section{Appendix title}

%If only one appendix, please use
%\appendix*
%\section{Appendix title}

%% After first section in \appendix*,
%% \section{} the term APPENDIX will not be used.

%=======================================================

%Use \bibliography{<name of your .bib file>}+
%to make your bibliography with BibTeX. 

%% to produce the bibliography,:
%%Make your bibliography by doing: pdflatex filename,  bibtex filename,
%% pdflatex filename,  pdflatex filename.

%%When uploading your files to Editorial Manager,  include both the .bib and the appropriate .bst file (for author/year reference style: jasaauthoryear2.bst; for numerical style: jasanum2.bst). Both the .bib and .bst should be uploaded as the “Manuscript (TeX or Word only)” item type.
\bibliography{sampbib}

%=======================================================

\end{document}